\documentclass[aps,pra,twocolumn,superscriptaddress]{revtex4-1}
\usepackage{graphicx}
\usepackage{hyperref}
\usepackage{amsmath}
\usepackage{subfigure}
\usepackage{epsfig} 
\usepackage{epstopdf}
\usepackage{float} 
\usepackage{tikz}
\usepackage[percent]{overpic}
\usepackage{comment}
\usepackage{pstricks}
\usepackage{placeins}
\usepackage{multirow}
\usepackage{rotating}
\usepackage{hhline}

\begin{document}
\newcommand{\Hm}{{\cal H}}
\newcommand{\lm}{\lambda_{\hbox{\scriptsize max}}}
\newcommand{\Uhat}{\hat{U}}

\title{Sensitivity to small perturbations in systems of large quantum spins}

\author{Tarek A. Elsayed}
\email{T.Elsayed@thphys.uni-heidelberg.de}
\address{Institute for Theoretical Physics, University of Heidelberg, Philosophenweg 19, 69120 Heidelberg, Germany}
\author{Boris V. Fine}
\email{B.Fine@thphys.uni-heidelberg.de}
\address{Institute for Theoretical Physics, University of Heidelberg, Philosophenweg 19, 69120 Heidelberg, Germany}
\address{Department of Physics, School of Science and Technology, Nazarbayev University,
53 Kabanbai Batyr Ave., Astana 010000, Kazakhstan}
\address{Skolkovo Institute of Science and Technology, 100 Novaya Str., Skolkovo, Moscow Region 143025, Russia}

\date{\today}

\begin{abstract}
We investigate the sensitivity of nonintegrable large-spin quantum lattices to small perturbations with a particular focus on the time reversal experiments known in statistical physics as ``Loschmidt echoes'' and in nuclear magnetic resonance (NMR) as ``magic echoes.''  Our numerical simulations of quantum spin-$7\frac{1}{2}$ clusters indicate that there is a regime, where Loschmidt echoes exhibit nearly exponential sensitivity to small perturbations with characteristic constant approximately equal to twice the value of the largest Lyapunov exponent of the corresponding classical spin clusters. The above theoretical results are verifiable by NMR experiments on solids containing large-spin nuclei.
\end{abstract}

\maketitle

\section{Introduction}

Exponential sensitivity to small perturbations is a defining property of classical chaos. 
Determining whether quantum systems exhibit the same property is important for understanding both quantum chaos as such and its role in the foundations of statistical mechanics. 

Many studies have been looking for static or dynamic manifestations of sensitivity to small perturbations in quantum systems that have chaotic classical limit \cite{chaudhury, benet,silvestrov,billiard,blumel}.  There are conflicting opinions about whether nonintegrable quantum systems can exhibit exponential sensitivity to perturbations. The often-mentioned  argument in favour of the absence of exponential sensitivity is that quantum mechanics is intrinsically linear, and, therefore, quantum amplitudes, which define all measurable properties, are not subject to nonlinear dynamics, which, in turn is required for the onset of classical chaos. Moreover, in bound quantum systems, the discreteness of the energy spectrum makes the evolution of any observable periodic or quasi-periodic in contrast to the  random behavior generated by the Lyapunov instability in the classical domain. A related argument is that the Heisenberg uncertainty relation does not allow one to define phase space trajectories, which, in turn, makes the notion of diverging phase space trajectories meaningless.

On the other hand, it is widely believed that the classical behavior should be restored from the quantum laws of motion in the limit of large quantum numbers. If  a quantum system has a chaotic classical limit, and if its initial wave function is a very narrow wavepacket, then the subsequent evolution of this wavepacket should follow the classical trajectory for a finite time interval and hence exhibit hypersensitivity to the perturbations of initial conditions.  The late-time recurrences exhibited by the wavefunction due to the discreteness of the energy levels do not contradict the possible initial instability.  This is similar to the situation in a bounded classical chaotic system where the local Lyapunov instability in the phase space does not prevent two initially diverging trajectories from coming arbitrary close to each other at a later time \cite{mendes}. For a macroscopic number of particles,  astronomical time scales will not be sufficient to observe the above recurrences - quantum or classical. From a somewhat different perspective, it can also be argued that the linearity of quantum mechanics should suppress sensitivity to small perturbations only as much as the linearity of Liouville's equation does for classical systems \cite{caves93}. More discussion of related issues can be found, for example, in Ref.~\cite{mendes}.

Particularly difficult in the above respect are nonintegrable macroscopic systems of spins 1/2, where macroscopic observables, such as the total magnetization, are expected to behave classically, but the individual microscopic constituents, i.e. spins 1/2, are as far from the classical limit as a quantum system can only be. 
It is natural in the context of this difficulty to adopt the following approach. One first substitutes spins 1/2 with classical spins and identifies the signatures of microscopic chaos in the behaviour of a macroscopic observable of the resulting classical system. Then one looks at the behavior of the same observable for the original spin-1/2 system to see whether it exhibits similar signature.

The task of identifying macroscopic signatures of microscopic chaos even for purely classical systems has been a long-standing challenge for the statistical physics community. One prominent approach was to try to extract the value of Kolmogorov-Sinai entropy (the sum of all positive Lyapunov exponents of the system) from the behaviour of a macroscopic observable \cite{gaspard-98}. This effort did not bring a conclusive outcome, because the true value of the Kolmogorov-Sinai entropy for macroscopic systems is not accessible either numerically or experimentally\cite{Dettmann-99,Schreiber-99,Cecconi-05,elsayed14}. Another approach was to look for the the exponential tails of relaxation functions\cite{Fine-03,Fine-04,Fine-05} and power spectra\cite{elsayed14} as signatures of microscopic chaos. It led to a number of experimentally verified predictions\cite{Morgan-08,Sorte-11,Meier-12}.  However, the difficulty of this approach is that, even though the above tails are indeed generic for chaotic many-particle systems, no quantitative connection is known between the exponential decay constants of these tails and the primary characteristics of chaos, namely, Lyapunov exponents of the system. 

Recently, we were able to make progress on the above agenda by first investigating the Lyapunov spectra of classical spin lattices\cite{deWijn-12,deWijn-13} and then identifying an experimentally feasible  manipulation of a classical spin system that would be able to access system's largest Lyapunov exponent\cite{fine14}. This manipulation is a weakly perturbed time reversal known in statistical physics as ``Loschmidt echo'' \cite{pastawski2000} and in nuclear magnetic resonance (NMR) as ``magic echo"\cite{MagicEcho, slichter}. The outcome of our analysis was that Loschmidt echoes in classical spin systems exhibit exponential sensitivity to small perturbations of perfect time reversal. The constant characterising this sensitivity is equal to twice the value of the largest Lyapunov exponent of the system. At the same time, we have shown that the corresponding spin-1/2 systems  are not exponentially sensitive to small perturbations even in the macroscopic limit, and, therefore, Lyapunov exponents cannot be defined for them. The above conclusions were supported by numerical simulations of Loschmidt echoes for spin-1/2 and for classical spin systems. 

The question then arises how the transition from non-exponential to exponential sensitivity of Loschmidt echoes proceeds. It is known\cite{largeN,lieb73,frohlich07} that the classical limit of quantum spin systems can be obtained by increasing the quantum spin number with the proper normalization of spin operators. As far as the Loschmidt echoes are concerned, our qualitative reasoning in Ref.~\cite{fine14} was that spin-1/2 systems are not exponentially sensitive to small perturbations, because one cannot slightly perturb an individual spin projection: any small perturbation creates a superposition of completely unperturbed state and a strongly perturbed state where the spin projection changes by 1, which means that the spin 1/2 flips entirely. From the above perspective,  a larger quantum spin $S$ has $2S+1$ possible values, which means that, if the spin projection is perturbed by 1, then there is still a range of values for the difference between unperturbed and perturbed evolution to grow before reaching the maximum value of the order of $2S$. Therefore, we expect that the range of nearly exponential sensitivity of Loschmidt echoes for systems of macroscopic number of spins $S$ to be of the order of $2S $. The goal of the present article is to verify numerically the presence of the above nearly exponential growth numerically.

\section{Numerical simulations of Loschmidt echoes}

\begin{figure} \setlength{\unitlength}{0.1cm}
\begin{picture}(88 , 107 )
{
\put(3, 55){ \epsfig{file=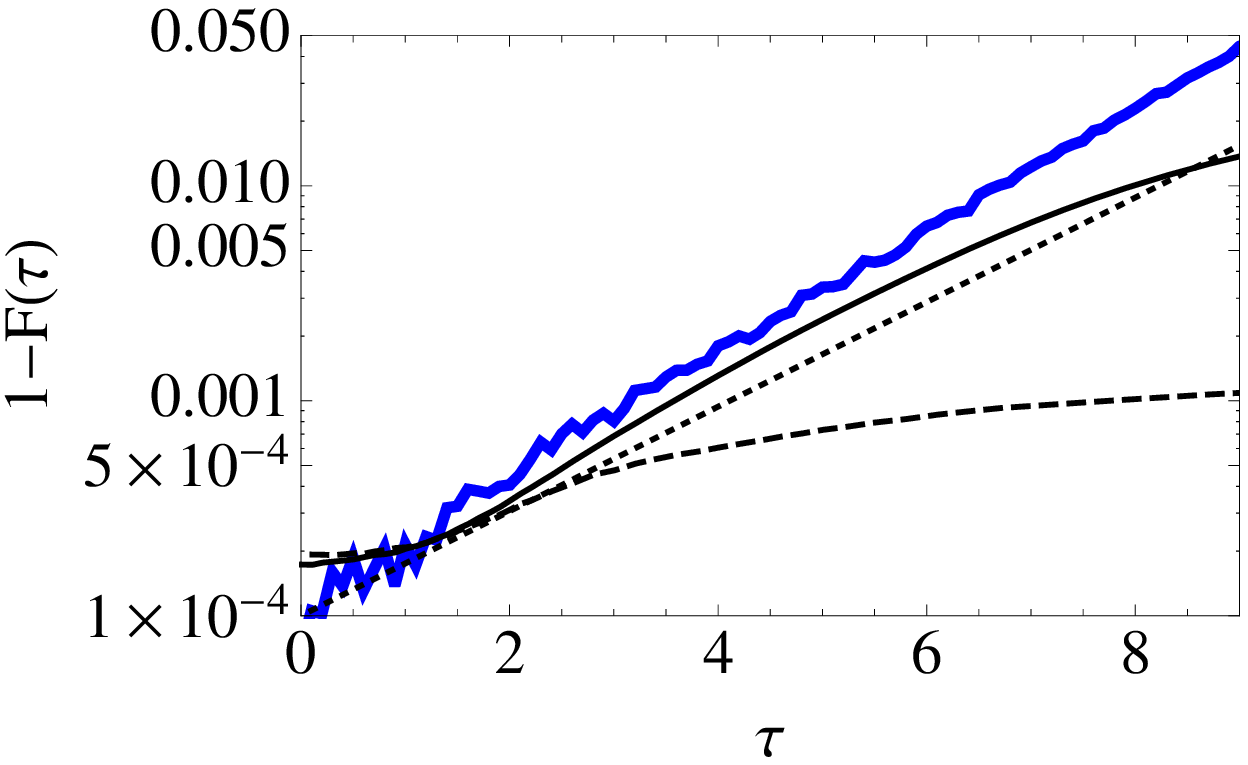,width=8cm } }
\put(3, 0){ \epsfig{file=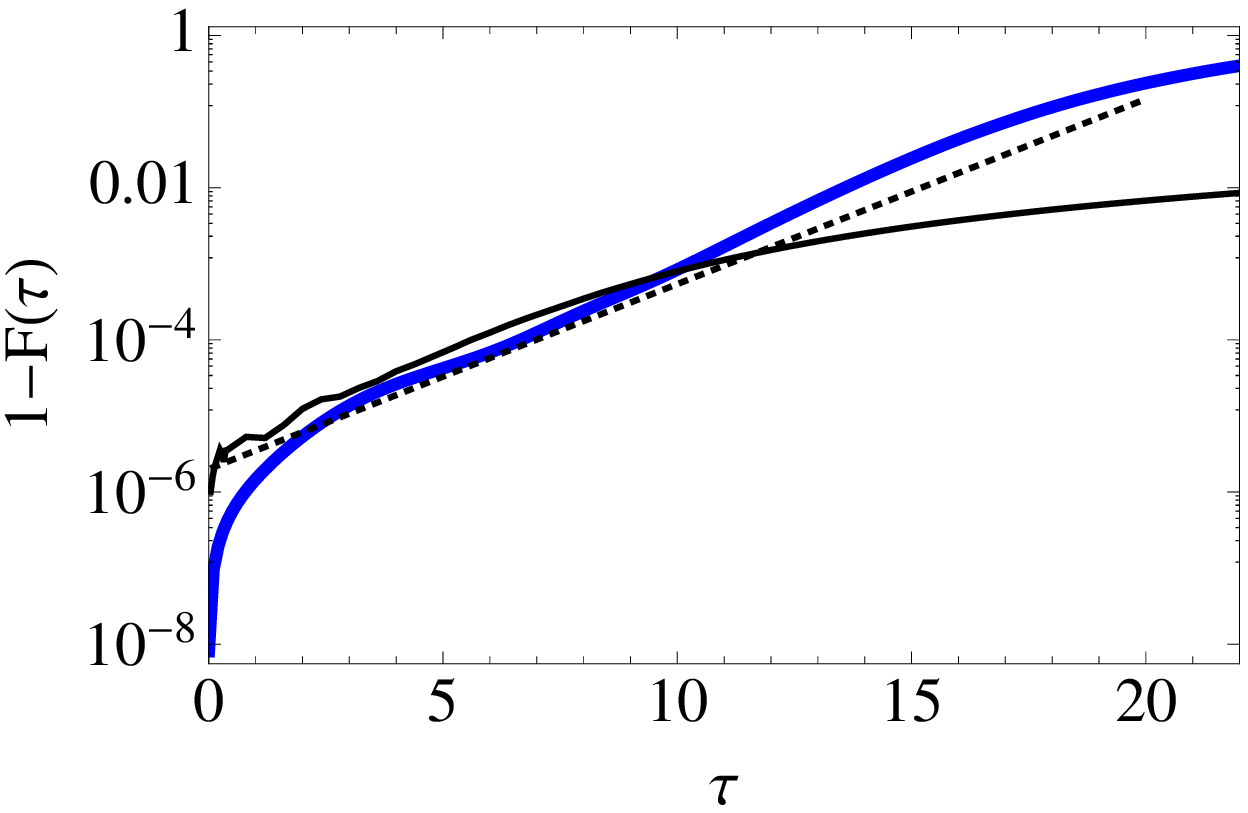,width=8cm } }
\put(2, 50){(b)}
\put(2, 100){(a)}
}
\end{picture}

\caption{ (Color online) Imperfect Loschmidt echoes $1-F(\tau)$. Thick solid blue lines represent  chain consisting of 6 classical spins as described in the text;  thin solid black lines represent the chain consisting of 6 quantum spins $7 {1 \over 2}$;  dotted lines represent the function $\alpha e^{2 \lambda_{max}}$, where $\alpha$ is a fitting parameter and $\lambda_{max}=0.28$ is the largest Lyapunov exponent of the classical system. (a) Loschmidt echo is perturbed at the moment of Hamiltonian reversal by rotating each spin around the $z$-axis by an angle randomly selected from the interval $[-\pi/100,\pi/100 ]$. For comparison, the dashed line represents  $1-F(\tau)$ for a quantum spin 1/2 chain consisting of 26 spins with interaction constants $J_z = 0.67$ and $J_{xy}= 0.33$. (b) Loschmidt echo is perturbed by adding a term $\Delta {\cal H} =\sum_k h_k S_{kz}$ to the reversed Hamiltonian, where each $h_k$ is randomly selected from the interval $[-0.002, 0.002]$. }
\label{qlyap}
\end{figure}

In the following, we compute Loschmidt echoes for quantum ($S=7{1 \over 2}$) and classical spin chains  consisting of 6 spins with periodic boundary conditions. The spins interact with the nearest-neighbor (NN) Hamiltonian
 \begin{equation}
{\cal H} =\sum_{m<n}^{\text{NN}}J_zS_m^z S_n^z+J_{xy} \left( S_m^x S_n^x+S_m^y S_n^y \right)
\label{H}
\end{equation}
where $S_i^{\alpha}$ represents either the quantum operator of the $\alpha^{\text{th}}$ ($x$, $y$ or $z$) projection of a quantum spin on $i^{\text{th}}$ lattice site or the corresponding projection of a vector of length 1 representing a classical spin. We take  $J_z=0.82$ and $J_{xy}=-0.41$ for the classical Hamiltonian, while for the quantum Hamiltonian we divide these values by the factor $\sqrt{S(S+1)}$ to match the characteristic timescale of the dynamics with the classical counterpart.  

Our Loschmidt echo manipulation is similar to NMR magic echo \cite{MagicEcho, slichter}. The system is initially slightly polarized in the $x$-direction in the vicinity of the infinite temperature equilibrium and then allowed to evolve under the action of the Hamiltonian (\ref{H}) for a certain time $\tau$. At time $\tau$, the sign of the Hamiltonian is reversed, and the system evolves under the action of the reversed Hamiltonian  for another time interval $\tau$. Afterwards, the value of the total magnetization in the $x$- direction  $M_x$ is registered.  

In real experiments, the reversal of the sign of the Hamiltonian is not perfect \cite{pastawski2000}. In our simulations, we intentionally introduce violations of the perfect time reversal by two methods:

(A) We apply small random rotations to the spins of the system at the moment of the Hamiltonian reversal. Specifically,  each spin is rotated around the $z$-axis by an angle randomly selected from the interval $[-\pi/100,\pi/100 ]$.

 (B) We add a small random-field perturbation $\Delta {\cal H} =\sum_k h_k S_{kz}$ to the Hamiltonian ${\cal H}$ during the backward evolution. The fields $h_k$ are randomly selected from the interval $[-0.002, 0.002]$.

We characterize the Loschmidt echo response by function 
\begin{equation}
F(\tau) = {\langle  M_x \rangle_f \over \langle  M_x \rangle_0},
\label{FTAU}
\end{equation}
 where $\langle \rangle_0$ and $\langle \rangle_f$ represent the  averages computed with respect to the probability distributions or density matrices for the initial and the final states of the system respectively.

It was shown in Ref. \cite{fine14} that for sufficiently small perturbations in chaotic classical spin systems, there is a regime of exponential departure from the perfect time reversal: 
\begin{equation}
1-F(\tau) \cong e^{ 2 \lm \tau},
\label{1mF}
\end{equation}  
where ${\lm}$ is the maximum Lyapunov exponent of the system. The above regime sets in after time $\tau$ of the order of $1/\lm$ required to suppress contributions from smaller Lyapunov exponents. The smallness of the initial perturbations should be such that $1-F(\tau) \ll 1$ in the regime exponential growth.  Eventually, as $\tau$ increases, the system leaves the regime of small perturbations and enters the saturation regime characterized by $1-F(\tau) \sim 1$. 

For quantum systems, the initial perturbation should be large enough to (ideally) make the overlap of perturbed and unperturbed wave functions equal to zero, but, at the same time, small enough to avoid entering the saturation regime immediately. The former restriction is of particular concern for finite-size numerical simulations:  together with the latter one, it leaves a rather limited growth range for the regime $1-F(\tau) \ll 1$.  For macroscopic systems, the former restriction is consistent with virtually any physically realizable small perturbations of either external fields or the interaction Hamiltonian. 

We computed $F(\tau)$ for the ring of 6 classical spins and the ring of 6 quantum spins $7 {1 \over 2}$ described above. In both cases, the initial nonequilibrium polarization along the $x$-axis was equal to 5\% of the maximum polarization.  For the classical system, we averaged over an ensemble of $7\times 10^6$ different initial conditions for both  perturbations (A) and (B). For the quantum system, it was sufficient, based on the recent results on quantum typicality\cite{elsayed2013,gemmer2009}, to use one initial quantum superposition state.  The equations of motion in both the classical and the quantum cases are propagated numerically by a fourth-order Runge-Kutta algorithm. In the classical case, $\lm$  was computed following the technique presented in Ref.\cite{elsayed14} and was found to be equal to 0.28. Both the calculations of $\lm$ and the classical simulations of Loschmidt echoes were done for zero total energy and zero total polarization along the $z$-axis. 

\section{Results and discussion}

In Fig. 1, we present the evolution of $1-F(\tau)$ for both the quantum and the classical spin systems for two types of perturbations (A) and (B), together with the fit of the form $\alpha e^{2 \lambda_{max} \tau}$, where $\alpha$ is a fitting parameter, and $\lm$ is the directly computed Lyapunov exponent. 

We observe that, in the quantum system, $1-F(\tau)$ exhibits a region of nearly exponential growth by about a factor of 10, which is consistent with our expectation that (i) this range exists, and (ii) the growth factor is limited by $2 S$.  We further observe that in the regime of nearly exponential growth for the spin-$7 {1 \over 2}$ chain, the growth rate is consistent with the rate for the corresponding classical spin system, which is, in turn, equal to $2 \lm$ with accuracy of about 15 percent. (This small discrepancy requires further investigation.) 

For comparison, we include in Fig. 1(a) the Loschmidt echo for a spin-1/2 chain. We define the term ``nearly exponential growth'' as the growth of $1-F(\tau)$  that can be well fitted by an exponential function over at least a factor of 10. With such a definition, the above chain of spins 1/2 exhibits no range of nearly exponential growth.

 The existence of the region of nearly exponential growth of  $1-F(\tau)$ for a large-spin quantum system  indicates that the experiments to determine effective Lyapunov exponents in materials with large quantum spins may be feasible.

\section{Conclusion}
We have shown numerically, to as much extent as  direct simulations allow, that systems of large quantum spins can produce signatures of exponential sensitivity to small perturbations similar to that of classical chaotic systems with positive Lyapunov exponents. This result is consistent with the expectation that classical behavior is reproduced by increasing spin quantum number $S$. It also indicates that, in solids containing large quantum spins, it is realistic to observe  a behavior similar to chaotic Lyapunov instabilities  with the help of NMR or other experimental techniques.  

 \bibliographystyle{apsrev4-1}

%

\end{document}